\documentclass[aps,prl,superscriptaddress,twocolumn]{revtex4-2}

\usepackage[english]{babel}
\usepackage[utf8x]{inputenc}
\usepackage[T1]{fontenc}
\usepackage{times}

\usepackage{subfigure}
\usepackage{t1enc,latexsym,amsfonts, amssymb, amsmath,empheq}
\usepackage{bm}
\usepackage{graphics}
\usepackage{graphicx}
\usepackage{url}

\usepackage[usenames,dvipsnames]{xcolor}

\usepackage{color}
\definecolor{darkblue}{rgb}{0,0,0.6}
\definecolor{darkred}{rgb}{0.6,0,0}
\definecolor{myblue}{RGB}{12, 12, 158}
\definecolor{myred}{RGB}{158, 19, 22}
\definecolor{myorange}{RGB}{245, 150, 12}
\definecolor{mygreen}{RGB}{26, 148, 49}
\definecolor{Prune}{RGB}{99,0,60}
\definecolor{Purple}{RGB}{75, 0, 130}
\definecolor{Pink}{RGB}{255, 105, 180}
\definecolor{deepskyblue}{RGB}{0, 191,255}
\definecolor{limegreen}{RGB}{50, 205, 50}
\definecolor{crimson}{rgb}{0.86, 0.08, 0.24}
\definecolor{coral}{rgb}{1.0, 0.5, 0.31}

\usepackage{hyperref}
\hypersetup{colorlinks=true,urlcolor=darkblue,citecolor=darkblue,linkcolor=MidnightBlue,hyperfootnotes=false}
\hypersetup{
bookmarksopen=true,
pdftitle="",
pdfauthor="Agoritsas-Bares", 
pdfsubject="", 
pdfstartview={FitH},		
pdfmenubar=true,			
pdfhighlight=/O,			
colorlinks=true,			
urlcolor=darkblue,
citecolor=darkblue,		
linkcolor=MidnightBlue,	
}
\hypersetup{colorlinks=true,urlcolor=darkblue,citecolor=darkblue,linkcolor=MidnightBlue,hyperfootnotes=false}

\newcommand{\arga}[1]{\left\lbrace #1\right\rbrace }

\newcommand{\valabs}[1]{\vert #1\vert}
\newcommand{\moy}[1]{\left\langle  #1 \right\rangle }

\def\de{\mathrm d}

\begin{document}


\title{Loss of memory of an elastic line on its way to limit cycles}

\author{Elisabeth Agoritsas}
\email[]{elisabeth.agoritsas@unige.ch}
\affiliation{Department of Quantum Matter Physics (DQMP), University of Geneva, Quai Ernest-Ansermet 24, CH-1211 Geneva, Switzerland}

\author{Jonathan Barés}
\email[]{jonathan.bares@umontpellier.fr}
\affiliation{Laboratoire de Mécanique et Génie Civil (LMGC), UMR 5508 CNRS-University Montpellier, 34095 Montpellier, France}

\date{\today}

\begin{abstract}

Oscillatory-driven amorphous materials forget their initial configuration and converge to limit cycles.
Here we investigate this memory loss under a non-quasistatic drive in a minimal model system,
with quenched disorder and memory encoded in a spatial pattern,
where oscillating protocols are formally replaced by a positive-velocity drive.
We consider an elastic line driven athermally in a quenched disorder
with bi-periodic boundary conditions and tunable system size,
thus controlling the area swept by the line per cycle
as would the oscillation amplitude.
The convergence to disorder-dependent limit cycles
is strongly coupled to the nature of its velocity dynamics depending on system size.
Based on the corresponding phase diagram, we propose a generic scenario for memory formation in driven disordered systems.

\end{abstract}


\maketitle

\paragraph{Introduction.}
\label{sec-intro}

Theoretical descriptions of driven amorphous materials remain challenging,
both for the rheology of yield stress fluids \cite{bonn_2017_RevModPhys89_035005}
or the mechanics of structural glasses \cite{arceri_landes_2020_Arxiv-2006.09725}.
%
Part of their complexity stems from their structural disorder
--self-generated by the relative position of their individual constituents--
which plays a key role in their response to an external drive.
%
%
There has been recently a collective endeavour to revisit the prominent features of these materials from the prism of \emph{memory formation} \cite{keim_2019_RevModPhys91_035002,nagel_2023_JChemPhys158_210401}.
%
Characterizing transient regimes,
towards steady-state or hysteretic behaviours,
amounts to study how materials forget an initial configuration
and eventually `learn' a new driving-dependent state.
The encoding and retrieval of such memories
in amorphous materials
is a promising avenue for metamaterials development
\cite{kadic_2019_NatureReviewsPhysics1_198,yu_2021_NatureReviewsMaterials6_226}.

In that respect, oscillatory protocols play a special role.
By tuning their amplitude, period, and spatial pattern, they allow us to systematically probe the explored disordered landscape.
Thermal activation is ill-controlled in comparison, probing energy barriers in a random and statistically isotropic way.
On the numerical side,
the oscillatory-athermal-quasistatic-shear (OAQS) protocols
provide bare characterizations of the landscape features with no characteristic timescale \cite{maloney_lemaitre_2006_PhysRevE74_016118}.
They have been the focus of several recent studies,
either in particle-based \cite{kawasaki_berthier_2016_PhysRevE94_022615,leishangthem_2017_NatureComm8_14653,adhikari_sastry_2018_EPJE41_105,yeh_2020_PhysRevLett124_225502,bhaumik_2021_PNAS118_e2100227118}
%
or coarse-grained models \cite{khirallah_2021_PRL126_218005,kumar_2022_JChemPhys157_174504,liu_2022_JChemPhys156_104902,parley_2022_PhysRevLett128_198001}.
%
Microscopically, an amorphous material experiences plastic events,
which locally update the structural disorder and generate a mechanical noise.
Upon this iterative restructuring,
and depending on the driving amplitude,
the material response departs from an initial elastic regime, and after one or several oscillatory cycles converges into a hysteretic behaviour.
Remarkably, the complex structure of these transient and limit-cycle responses upon OAQS can further be revealed
via random transition graphs between mechanically stable configurations \cite{mungan_witten_2019_PhysRevE99_052132,mungan_2019_PhysRevLett123_178002,regev_2021_PhysRevE103_062614,mungan_2021_PhysRevLett127_248002}.
%

On the experimental side, however, oscillatory protocols are often performed at \emph{finite} driving rate, typically with a sinusoidal shear strain (or stress),
whose frequency dependence is central in rheological measurements of storage and loss moduli \cite{bonn_2017_RevModPhys89_035005,wang_prl_2020,xing_prl_2021,keim_sa_2021,zhao_prx_2022}.
The additional timescale introduced by the drive compels us to clarify the interplay between elastic relaxation and disorder dynamics.
In particular, it questions the generalisability of the memory picture built on quasistatic protocols to actual rheological experiments.

In this Letter,
we investigate numerically a minimal model system, devised to disentangle
these memory effects from the structural disorder dynamics,
focusing on displacement-controlled protocols at \emph{finite} velocity.
It aims to bypass three
main difficulties:
first, in amorphous materials,
configurations and disorder are parametrized by the same degrees of freedom, namely the positions of individual particles;
secondly, we are better equipped theoretically to address overdamped steady states at positive driving, rather than hysteretic behaviours under oscillating protocols;
thirdly, a bidimensional system permits direct visualization of what happens, especially for the memory process, and faster simulations.
Our model system consists of an elastic line
driven athermally
in a bi-periodic 2D random landscape
by adjusting the center-of-mass velocity.
The disorder is quenched, with \textit{ad hoc} features that we fully control.
Memory is encoded in the geometrical \emph{and} velocity profiles of the line.
And oscillating protocols are formally replaced by a positive velocity drive:
instead of driving the line back and forth swiping repeatedly the same landscape, the line is continuously driven over a repeated landscape.
Hence,
tuning the system size controls the area swept by the line per cycle, as would an oscillatory amplitude.

Our findings put forward the key role of velocity dynamics regarding memory formation.
%
In the absence of thermal fluctuations, the line converges to disorder-dependent limit cycles.
In quasistatic,
the limit cycle is then fully characterized by the sole shape of the line,
optimized in the successive minima of the disordered landscape.
For a finite driving rate,
on the contrary, part of the memory is encoded in the velocity profile.
Therefore,
the way the line forgets its initial conditions will also strongly depend on the nature of the velocity dynamics,
self-consistently fixed by the disorder/size/driving settings.

\paragraph{Model.}
\label{sec-model}

\begin{figure}[th]
\begin{center}
\includegraphics[width=0.95\columnwidth]{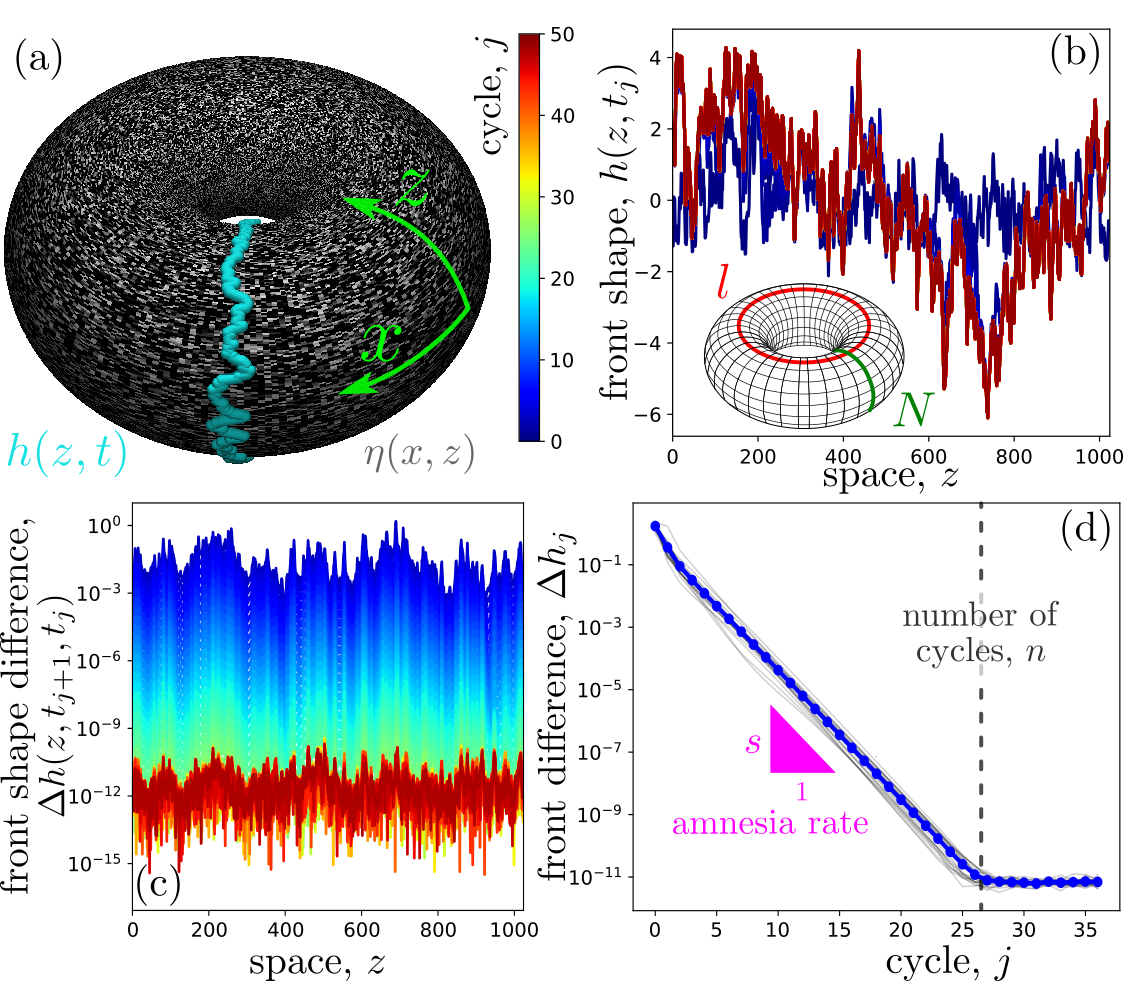}
\caption{
\textbf{Settings}:
{(a)}~Quenched disorder ${\eta(x,z)}$
embedded in a bi-periodic 2D space of coordinates ${(x,z)\in [0,\ell] \times [0,N]}$;
driven elastic line parametrised 
by the univalued profile ${h(z,t)}$.
{(b)}~Stroboscopic view of its geometrical profile when ${\bar{h}(t_j)=0}$.
{(c)}~Profile difference, in absolute value, between successive cycles. 
{(d)}~Exponential decay of the integrated profile difference ${\Delta h_j}$,
plateauing after $n$ cycles.
Parameters:
$c=10^{-3}$, $k=10^{-2}$, $\sigma=1$, $N=1024$, $l=10$.
}
\label{fig-schema-model}
\end{center}
\end{figure}

We consider an elastic line evolving in a 2D space of coordinates ${(x,z) \in [0,\ell] \times [0, N]}$ with bi-periodic boundary conditions,
as shown in Fig.~\ref{fig-schema-model}(a).
In the absence of overhangs,
this front shape is parametrized by an univalued profile $h(z,t)$.
It evolves in a quenched random field ${\eta(x,z)}$,
with the overdamped dynamics
\begin{equation}
\label{eq-Langevin-dynamics}
\begin{split}
& \partial_t h(z,t)
= 	\underbrace{c t - k \moy{h(z,t)}_z}_{\text{speed loading, $F_{\rm{load}}$}}
	+ \underbrace{F_{\text{st}}[h(z,t)]}_{\text{stiffness}}
	+ \underbrace{\sigma \, \eta(h(z,t),z)}_{\text{disorder}}
\\
& \text{with} \quad F_{\text{st}}[h(z,t)]=
\frac{1}{\pi} \int_0^N \!\!\!\! \de z' \, \frac{h(z',t)-h(z,t)}{\valabs{z'-z}^\gamma}
\end{split}
\end{equation}
with the loading rate $c$,
unloading factor $k$,
and disorder amplitude $\sigma$.
${\moy{\bullet}_z}$ denotes the spatial average in the $z$ direction,
hence ${\moy{h(z,t)}_z \equiv \bar{h}(t)}$ is the center-of-mass.
We assume a Gaussian disorder of zero means:
denoting $\overline{\bullet}$ the average over disorder realisations,
$\overline{\eta(x,z)}=0$
and
$\overline{\eta(x,z)\eta(x',z')}=2 \, \delta (x-x') \delta (z-z')$
\textit{i.e.}~spatially uncorrelated above the numerical discretization scale.
The `stiffness' elasticity range can be tuned by changing the exponent $\gamma$,
and to focus for now on a single model here we set ${\gamma=2}$.
The remaining parameters are tuned to probe the different amnesia regimes :
$c \in [10^{-5},5 \times 10^{-2}]$,
$k \in [10^{-3},5 \times 10^{-1}]$,
$\sigma \in [0.1,5]$,
and in pixel-size units
$N \in [128,16384]$
and $l \in [2,400]$.

This is one representative example of the theoretical framework of \emph{disordered elastic systems},
successfully applied to a broad range of physical interfaces (ferroic domain walls, imbibition, proliferating cells fronts, \textit{etc.})
to address the role of disorder in their properties
\cite{agoritsas_2012_ECRYS2011,wiese_2022_RepProgPhys85_086502,fisher_1998_PhysReports301_113,brazovskii_2004_AdvPhys53_177,santucci_2011_EurPhysLett94_46005}.
%
We focus on the long-range elastic line ($\gamma=2$) with `random-field' disorder,
extensively investigated as a paradigm of brittle cracks in fracture mechanics \cite{tanguy_1998_PhysRevE58_1577,maloy_2006_PhysRevLett96_045501,
alava_2006_AdvPhys55_349,
ponson_2010_InternationalJFracture162_21,
bares_2013_PhysRevLett111_054301,
bares_2018_NatureComm9_1253,
bares_2019_PhilTransRSocA377_20170386,
bares_2019_PhysRevE100_023001}.
It is also relevant for driven amorphous materials, as it retains the long-range nature of the Eshelby stress propagator, albeit without its non-convex features \cite{lin_2014_PNAS111_14382,tyukodi_2016_PhysRevE93_063005}. Moreover, our driving is formally closer to a genuine mechanical loading.
%
Our model's critical features in the steady state are well-benchmarked within the depinning formalism \cite{ertas_kardar_1994_PhysRevE49_R2532,demery_rosso_ponson_2014_EurPhysLett105_34003,
demery_2014_JStatMech2014_P03009,bares_2013_PhysRevLett111_054301,bares_2021_PhysRevE103_053001},
allowing us to focus directly on memory issues. 
%
In a nutshell:
\textit{(i)}~The speed loading $F_{\text{load}}(t)$ is a competition between an external loading rate $c$ and a restoring stiffness $k$ acting on the center-of-mass.
This displacement-controlled driving guarantees that the line will always reach a finite steady velocity and ${F_{\text{load}}^{\text{steady}} (t)\geq F_c}$,
where $F_c$ is a critical force set by the disorder configuration \cite{bolech_rosso_2004_PhysRevLett93_125701}.
%
\textit{(ii)}~Conversely, in an athermal constant force driving, $F_c$ is the minimal force needed to reach such a self-sustained steady state.
%
%
\textit{(iii)}~The nature of the velocity dynamics changes radically
depending on the proximity to $F_c$:
in the `fast-flow' or continuous regime,
the average velocity displays small fluctuations around the mean $c/k$,
\textit{i.e.}~${F_{\text{load}}^{\text{steady}}(t) \approx c/k \gg F_c}$;
closer to $F_c$,
the line advances intermittently via critical avalanches,
in the so-called `crackling' regime \cite{bares_2013_PhysRevLett111_054301}.
%
%
These features 
are generic for many driven disordered systems,
and 
reminiscent of the yielding transition of amorphous materials \cite{baret_vandembroucq_2002_PhysRevLett89_195506,lin_2014_PNAS111_14382,tyukodi_2016_PhysRevE93_063005}.
Regarding the present study,
we could have considered alternative choices for the elasticity and disorder, similarly benchmarked.
Yet, with our choice 
we anticipate reintroducing disorder dynamics in our model system:
\cite{bares_2021_PhysRevE103_053001}
provides an additional benchmark on how artificial disorder updates trigger avalanches and modify their distributions in a controlled way.

\paragraph{Memory characterisation.}
\label{sec-memory-characterisation}

The transverse size $\ell$ is usually considered as a necessary nuisance in numerical simulations \cite{bustingorry_2010_PhysRevB_82_094202},
whereas here it is precisely our control parameter of interest.
%
We start with a flat and still initial condition (${h(z,0)=\partial_t h(z,0) = 0 \, \forall z}$),
and drive deterministically the line with its overdamped dynamics
over $j$ cycles in the $x$-direction.
We take a stroboscopic snapshot of its shape 
each time $t_j$ its center-of-mass crosses ${x=0}$.
We then quantify the convergence to a limit cycle, for a given disorder realization,
by comparing consecutive profiles
${h_j(z)\equiv h(z,t_j)}$.
%
These are typically quite close (see Fig.~\ref{fig-schema-model}(b)),
thus we focus instead on their front shape difference
\begin{equation}
\label{eq-front-shape-difference}
\begin{split}
& \Delta h(z,t_{j+1},t_{j})
\equiv \left| h(z,t_{j+1}) - h(z,t_{j}) \right| ,
\\
& \Delta h_j
	\equiv \moy{\Delta h(z,t_{j+1},t_{j})}_z
	= \frac{1}{N} \int_0^N \!\!\! \de z \, \Delta h(z,t_{j+1},t_{j}) ,
\end{split}
\end{equation}
as shown in Fig.~\ref{fig-schema-model}(c)-(d).
%
By plotting the average difference as a function of the cycles,
we can follow the memory loss and convergence to a limit cycle.
See Supplementary Information (SI) for alternative criteria,
including the velocity profiles ${v_j(z)\equiv \partial_t h(z,t_j)}$,
supporting that we can safely focus on the most straightforward indicator, $\Delta h_j$.

As shown in Fig.~\ref{fig-schema-model}(d),
the qualitative trend is an exponential decrease of ${\Delta h_j}$
until it plateaus.
We aim to understand what controls the number of cycles $n$ needed to reach the plateau $\Delta h_\infty$.
The latter would be strictly zero if the line was reaching a unique limit cycle, as expected in an athermal quasistatic driving by Middleton's theorem \cite{middleton_1992_PhysRevLett68_670}
assuming a perfect numerical resolution.
%
In practice, the key quantity to extract is the average slope of ${\Delta h_j}$ in semi-log scale, \textit{i.e.} the amnesia rate ${s \approx - \partial_j \log \Delta h_j}$.
For an exponential decrease ${\Delta h_{(j \leq n)} = A e^{-s j}}$
and a negligible plateau,
we have ${n= - \log (\Delta h_\infty/A)/s \approx \log (A) /s}$
where $A$ depends on the initial condition.
The inverse amnesia rate is a good proxy for the number of cycles (${n \sim s^{-1}}$), much less sensitive to initial conditions and independent of
the plateau value (see SI).
The figures reported here are based on this quantity.

\paragraph{Number of cycles for amnesia.}
\label{sec-cycles-for-amnesia}

\begin{figure}[h]
\begin{center}
\subfigure{\includegraphics[width=0.97 \columnwidth]{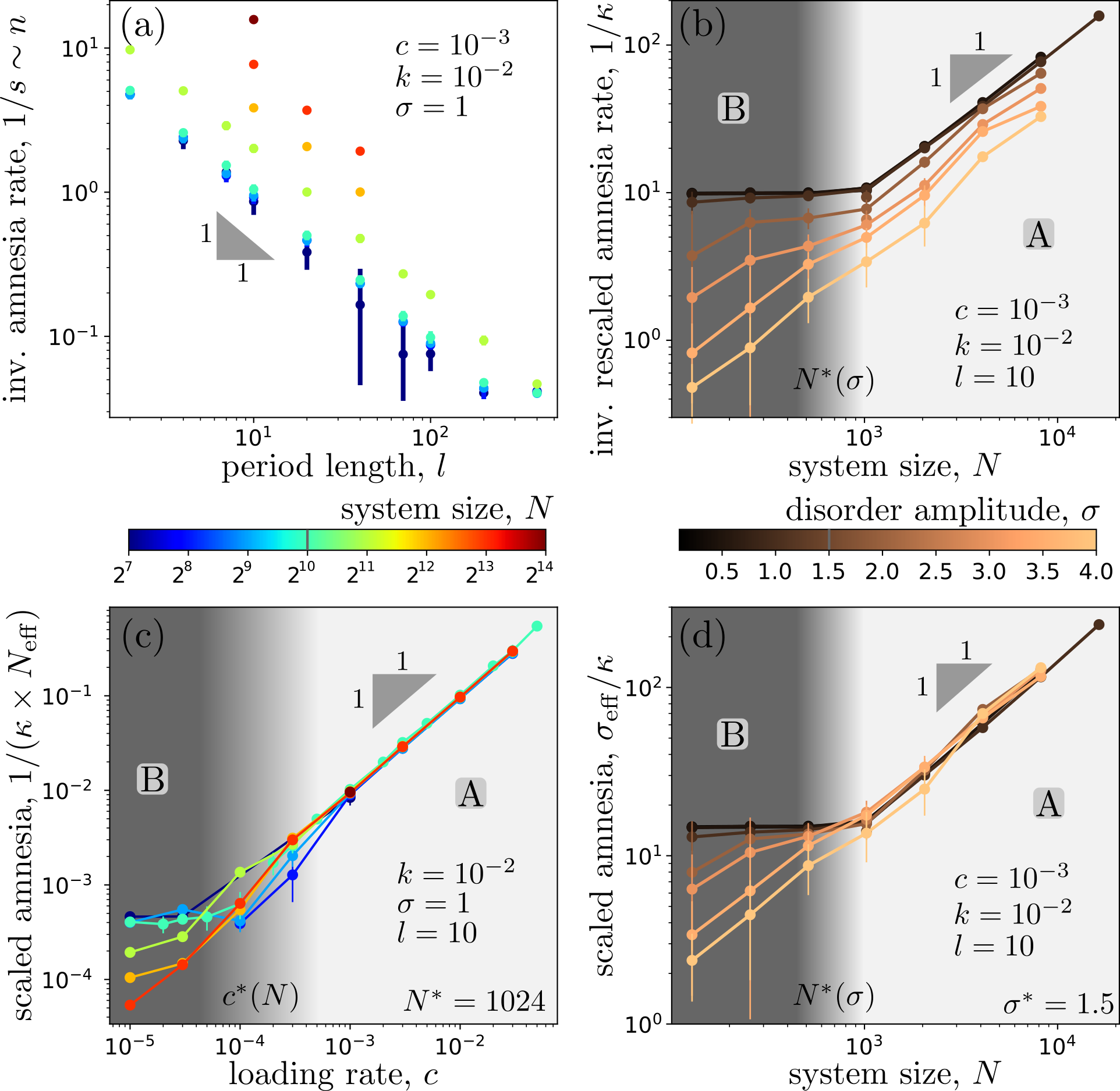}}
\caption{
\textbf{Quantifying amnesia}:
{(a)}~Inverse amnesia rate $1/s \sim n$
as a function of transverse size $\ell$, for different system size $N$.
{(b)}~Proxy for the distance to amnesia, ${1/\kappa}$, as a function of $N$, for different disorder amplitude $\sigma$.
For small $\sigma$, two distinct regimes separated by ${N^*(\sigma)}$ are highlighted with grey shading (\textbf{A}-right, \textbf{B}-left).
{(c)}~Inverse amnesia rates as a function of the loading rate $c$, for different system size $N$.
Collapse in \textbf{A} is obtained upon rescaling with ${N_{\text{eff}}(N>N^*)=N}$ and ${N_{\text{eff}} \approx N^*}$ otherwise; this reveals a crossover at loading $c^*(N)$ higlighted in grey shading.
(d)~Partial collapse in \textbf{A} of (b) upon rescaling with $\sigma_{\text{eff}}(\sigma>\sigma^*)=\sigma$ and ${\sigma_{\text{eff}} \approx \sigma^*}$ otherwise.
Error bars are for disorder fluctuations.
}
\label{fig-crossover-amnesia}
\end{center}
\end{figure}

For the considered parameter range, the plateau value ${\Delta h_\infty} \approx 10^{-13}$ is fixed by the numerical resolution and thus negligible.
Keeping the parameters $\arga{c,k,\sigma}$ fixed,
we find that $n$ decreases linearly in the transverse size $\ell$ (see Fig.~\ref{fig-crossover-amnesia}(a)).
This suggests to consider instead ${d_n=n\ell}$, \textit{i.e.} the distance covered by the center-of-mass 
to reach amnesia,
and we focus on the \emph{rescaled amnesia rate} defined as $\kappa = s/\ell \sim d_n^{-1}$.
By plotting this proxy for $d_n$ with respect to $N$
(see Fig.~\ref{fig-crossover-amnesia}(b)),
two distinct regimes arise for the small $\sigma$:
linear above a given $N^*(\sigma)$ (\textbf{A}, pale region)
and constant below (\textbf{B}, dark region).
Above $N^*$, the number of cycles for amnesia is proportional to the aspect ratio of the system size: ${\kappa^{-1} \sim d_n \sim N}$ thus $n \sim N/\ell$.
In \textbf{A}, memory is extensive.
%
Below $N^*$, the distance $d_n$ saturates at ${d_n \sim N^*(\sigma)}$.
In \textbf{B}, there is a minimal distance required to forget, which can only depend on the remaining parameters.
Under a quasistatic drive $d_n$ must be fixed by the landscape features ($\sigma$),
whereas at finite driving it may also depend on how fast the line is driven (${c,k}$).

\paragraph{Amnesia rate v.s. loading and disorder.}
\label{sec-rescaled-amnesia-rate}

To further understand how the memory extensivity is broken,
we thus vary independently $\arga{c,\sigma}$, while imposing ${k=0.01}$ without any loss of generality.
In Fig.~\ref{fig-crossover-amnesia}(b), regimes \textbf{A} and \textbf{B} were clearly distinct for small $\sigma$ on which the grey shading is based,
whereas upon increasing disorder the data spread and render the saturation towards \textbf{B} less apparent in log scale.
We thus focus on comparing curves where memory is extensive and rescale them such as to collapse their \textbf{A}~regime.
Fig.~\ref{fig-crossover-amnesia}~(c)-(d) suggests the existence of characteristic $c^*$ and ${1/\sigma^*}$, respectively,
below which the amnesia regime departs from ${\kappa^{-1} \sim c / \sigma}$.

These linear scalings and crossovers are not to be taken as robust characterizations, they merely highlight the general trend:
memory is extensive as long as the disorder is weak (below $\sigma^*$), the system is large (above $N^*$) or the driving is fast (above $c^*$).
In \textbf{A} the number of cycles ${n \approx s^{-1} \sim (N/\ell) (c/\sigma)}$
implies in particular that it takes fewer cycles to converge to the limit cycle
at lower loading.
We eventually depart from this regime at ${c<c^*(N,\ell,\sigma)}$, which includes the quasistatic limit.
This strongly suggests examining the velocity dynamics associated with these different behaviours.

\paragraph{Relevance of the velocity dynamics.}
\label{sec-memory-velocity}

\begin{figure}[!htb]
\begin{center}
\subfigure{\includegraphics[width=0.99\columnwidth]{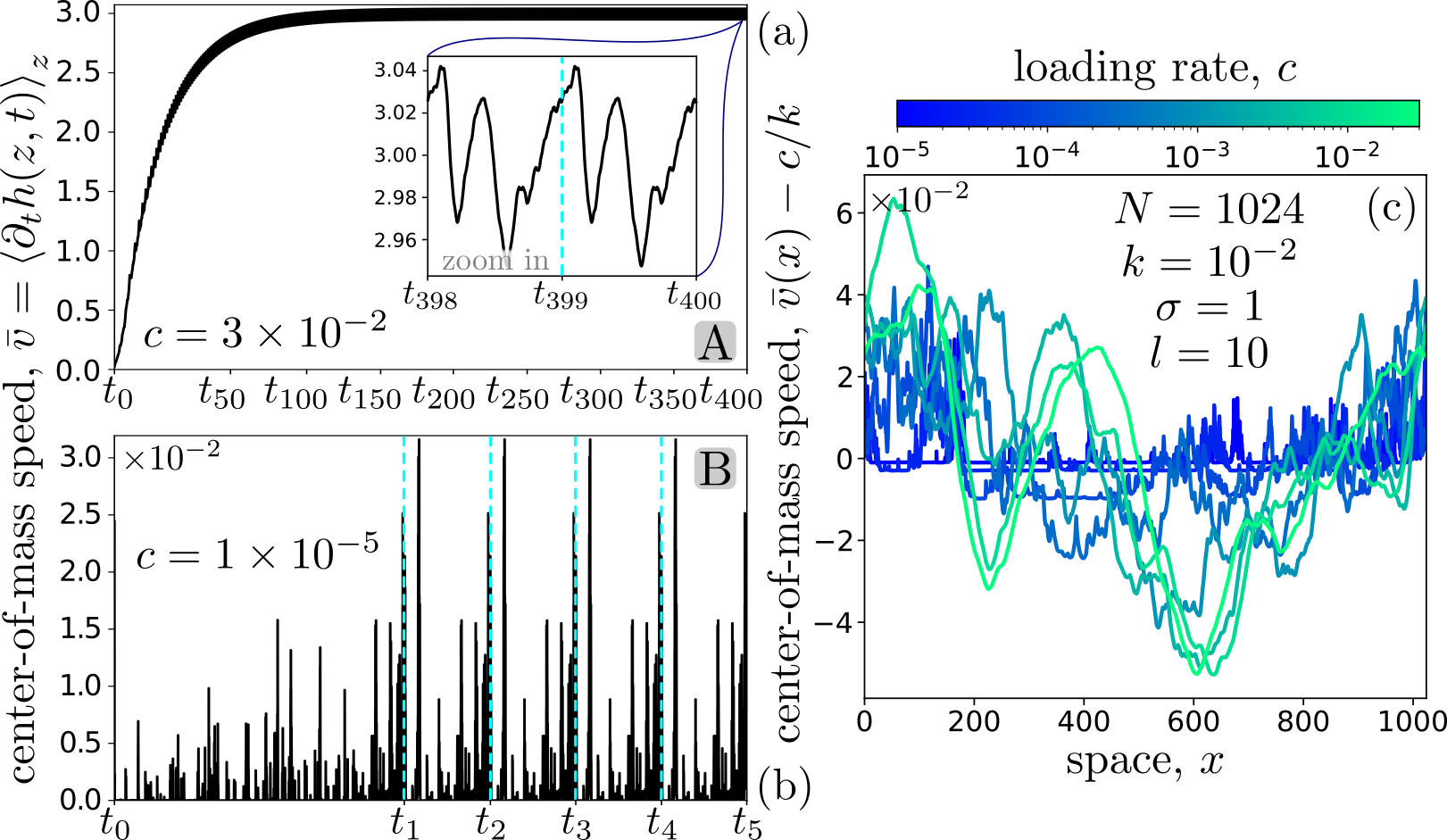}}
\caption{
\textbf{Center-of-mass dynamics}:
(a)-(b)~Evolution of the center-of-mass speed $\bar{v}(t)$ for fast (\textbf{A}) and slow (\textbf{B}) loading,
matching the two regimes in Fig.~\ref{fig-crossover-amnesia}(a).
Times $\arga{t_j}$ when the system begins a new cycle are given on the horizontal axis.
(c)~Center-of-mass speed $\bar{v}(x)$ along a cycle in the converged regime for increasing speed rate, for the same disorder and initial condition. Data are shifted by the average speed $c/k$, and mapped on the spatial interval ${x(t) \in [0,\ell]}$.
}
\label{fig-velocity-centerofmass}
\end{center}
\end{figure}

The velocity profiles also converge to limit cycles (see SI),
yet the center-of-mass speed $\bar{v}(t)$
is more informative on what happens \emph{within} each cycle.
Fig.~\ref{fig-velocity-centerofmass}(\textit{left}) illustrates two very distinct behaviours,
as expected from previous studies \cite{bares_2013_PhysRevLett111_054301}:
\textbf{A} converges to a continuous regime in the limit cycle,
and \textbf{B} to an intermittent behaviour.
The only difference between these two cases is the loading rate, with \textbf{A} at ${c > c^*(N)}$
and \textbf{B} otherwise.
In Fig.~\ref{fig-velocity-centerofmass}(c)
we further show how varying the loading rate modifies the limit cycle for ${\bar{v}(t)}$,
for a given disordered landscape.

%
These findings support the following scenario:
For a given landscape of parameters ${\arga{N,\ell,\sigma}}$,
the speed loading ${\arga{c,k}}$
determines if the dynamics in the limit cycle are either continuous ($\bar{v} \approx c/k$ with small fluctuations) or intermittent (but always strictly positive).
The associated transient regime from flat initial conditions will share similar qualitative features.
In the continuous case (\textbf{A})
the line is driven so fast that it flies over the landscape, without fully relaxing in its minima;
it consequently needs more cycles to learn its features,
its amnesia rate is smaller,
and memory is extensive (${d_n \sim N}$).
In the intermittent case (\textbf{B}),
the line is driven so slowly that it follows closely the landscape;
disorder features are encoded more efficiently \textit{i.e.} in less cycles,
amnesia rate is larger,
and extensivity is broken.

\paragraph{Unifying phase diagram.}
\label{sec-phase-diagramme}

\begin{figure}[h]
\begin{center}
\subfigure{\includegraphics[width=0.90\columnwidth]{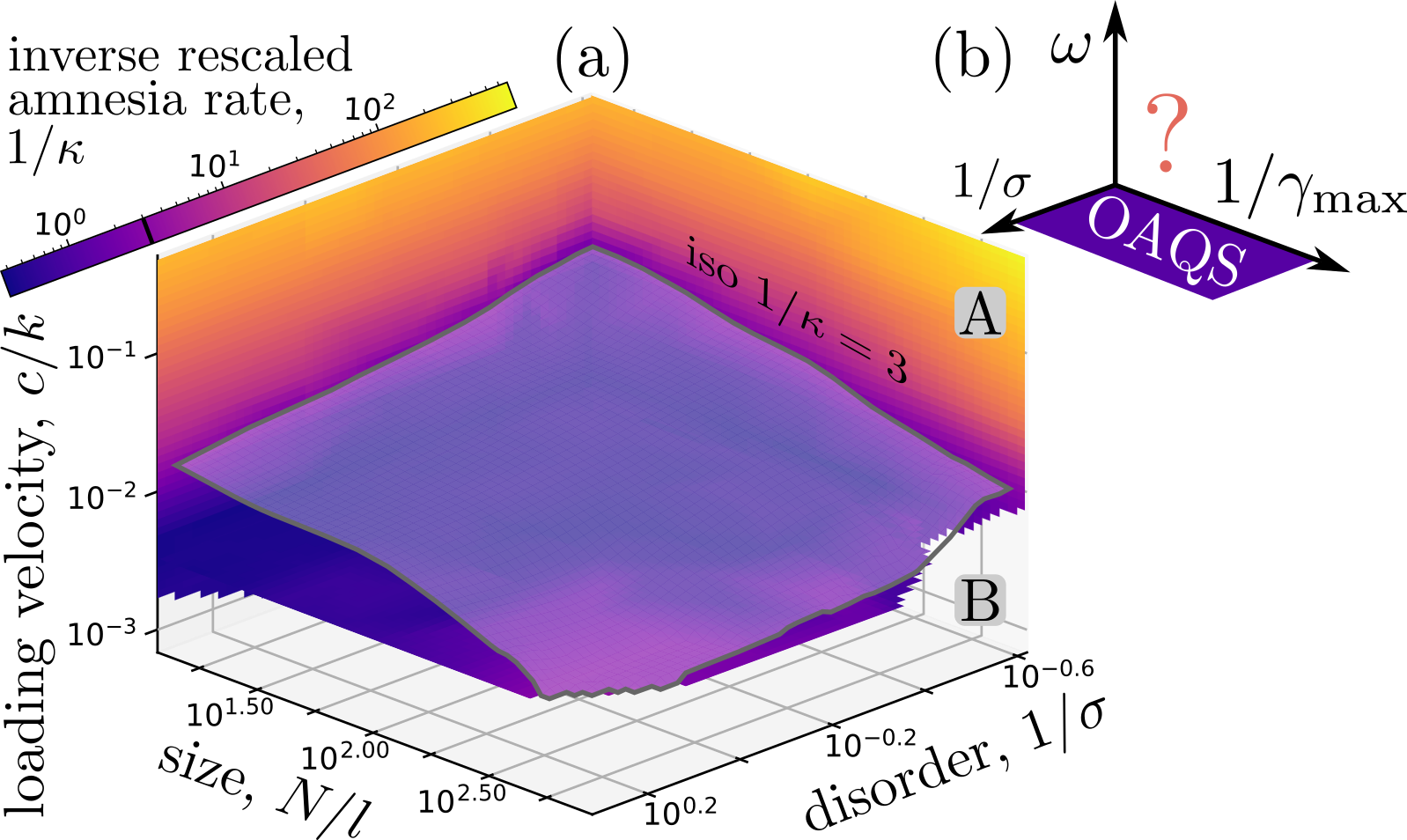}}
\caption{
\textbf{Phase diagramme of memory formation \textit{versus}  nature of the velocity dynamics}:
continuous (\textbf{A}) above the iso-surface and intermittent (\textbf{B}) below.
The iso-surface ${\kappa=1/3}$ is an arbitrary threshold, below which $\kappa$ mainly saturates.
See SI for an animated plot.
}
\label{fig-phase-diagramme}
\end{center}
\end{figure}

For such disordered elastic systems,
the crossover between these two types of behaviours is highly non-trivial for steady states \cite{wiese_2022_RepProgPhys85_086502, bares_2013_PhysRevLett111_054301,bares_2019_PhysRevE100_023001},
and \textit{a fortiori} for transient regimes.
To support our scenario,
we plot in Fig.~\ref{fig-phase-diagramme}
the rescaled amnesia rates ${\kappa \sim 1/(n\ell)}$ 
through the 3D phase space of our minimal model.
The three axes control
the aspect ratio ${N/\ell}$,
the inverse amplitude of disorder,
and the forcing $c/k$. 
For each point, we can discriminates between continuous (\textbf{A}) or intermittent (\textbf{B}) transient behaviours,
albeit with a quantitative dependence on the thresholding (see SI).
It is consistent with the saturation of $\kappa$ 
highlighted by the iso-surface ${\kappa=1/3}$,
confirming our physical picture.
This phase diagram thus rationalizes the memory formation at \emph{finite} velocity, in a tunable disordered landscape.

\paragraph{Insights for driven amorphous materials.}
\label{sec-implications-amorphous}

A genuine oscillatory protocol at finite velocity amounts to replace
in Eq.~\eqref{eq-Langevin-dynamics} the loading by a sinusoidal force, of amplitude ${\gamma_{\text{max}}}$ and frequency $\omega$ \cite{brazovskii_2004_AdvPhys53_177,nattermann_2001_PhysRevLett87_197005,glatz_nattermann_pokrovsky_2003_PhysRevLett90_047201,schuetze_2010_PhysRevE81_051128, schuetze_nattermann_2011_PhysRevB83_024412}.
In our model, ${\gamma_{\text{max}}}$ translates formally into the aspect ratio $\ell/N$, and $\omega$ into the loading ${v=c/k}$.
According to our phase diagram,
increasing $\gamma_{\text{max}}$ or decreasing $\omega$ is expected to radically change the memory formation, making it more efficient (fewer cycles),
with a crossover from continuous to intermittent dynamics \cite{zadeh_2019_PhysRevE99_040901}.
The precise location in this diagram
will further depend on the disorder strength $\sigma$.
For amorphous materials, the latter is self-consistently determined for a given driving protocol, 
and \textit{a priori} not constant in transient regimes.
%
This implies that, in order to rationalize memory formation under non-quasistatic oscillating protocols,
one would crucially need to characterize jointly 
the driving-dependent disorder strength and the resulting nature of the velocity dynamics, at fixed ${\arga{\omega,\gamma_{\text{max}}}}$.

The well-studied OAQS regime corresponds to a plane of our putative phase diagram,
as sketched in Fig.~\ref{fig-phase-diagramme}.
Adding a finite driving rate is bound to unveil a more complex behaviour,
as already displayed by our minimal model with frozen disorder dynamics.
Here we showed that driven disordered systems can display very different memory behavior,
and traced this back to how the velocity dynamics is self-consistently fixed by the disorder/size/driving settings.
This is the core of the depinning-like phenomenology, hence this scenario should hold more broadly for disordered elastic systems.
%
By construction, 
our specific model system is not expected to reproduce the full phenomenology of oscillatory-driven amorphous materials; 
yet, the analogy between the depinning and yielding transitions in coarse-grained models \cite{baret_vandembroucq_2002_PhysRevLett89_195506,lin_2014_PNAS111_14382,tyukodi_2016_PhysRevE93_063005}
strongly suggests challenging our generic scenario already at this level. 

\paragraph{Concluding remarks.}
\label{section-conclusion}

We started from a flat configuration and examined how the line converges to its limit cycle jointly for its geometric and velocity profiles.
Focusing on the rescaled amnesia rate $\kappa$, we find that there is a regime where the memory is extensive,
\textit{i.e.} $\kappa^{-1}$ is proportional to the system size $N$, the loading and the inverse of the disorder amplitude.
It requires more cycles to `learn' the quenched landscape when the latter is small or the line is driven faster.
If we start from a configuration closer to the limit cycle, this number of cycles will be reduced accordingly.
On the contrary, the amnesia rate will not be affected,
promoting it to the key physical quantity to follow in further studies,
and in particular to investigate the intermittent regime where $\kappa$ saturates.

The scenario we put forward is generic for an extended object driven repeatedly over a given patch of disordered landscape,
until the geometry of this object fully encodes the finite amount of information associated with the disorder.
The specific scalings will of course vary on the microscopic model, for instance
we expect the memory extensivity to generalize to ${n \sim N/\ell^{f(\gamma)}}$
when tuning the elasticity range.
In this regime, the limit cycles correspond to the so-called fast-flow 
regime in a `random-periodic' disorder settings \cite{bustingorry_2010_PhysRevB_82_094202}, thus
paving the way for an analytical study of transient regimes \cite{agoritsas_bares_2023_long}.

Finally,
our minimal model system provides a controlled framework to systematically investigate the joint role of initial conditions and disorder features in memory formation.
In addition to allowing for a finite spatial correlation length \cite{agoritsas_2012_FHHtri-analytics,agoritsas_garcia-garcia_2016_JStatPhys164_1394} and/or a more complex hierarchical structure of the landscape \cite{charbonneau_2014_NatureCommunications5_3725},
two key ingredients to address in the future are tunable disorder dynamics as in \cite{bares_2021_PhysRevE103_053001}
and different initial conditions.
Both could be tailored to mimic the self-consistent disorder dynamics in driven amorphous materials, in order to probe its relevance for memory in finite driving protocols \cite{regev_2013_PhysRevE88_062401,paulsen_2014_PhysRevLett113_068301}.


\begin{acknowledgments}

We thank Muhittin Mungan, Alberto Rosso, and Damien Vandembroucq for fruitful discussions.
E.A. acknowledges support from the Swiss National Science Foundation by the SNSF Ambizione Grant No.~PZ00P2{\_}173962, and from the Simons Foundation (Grant No.~348126 to Sidney Nagel).

\end{acknowledgments}



\bibliographystyle{plain_url}

\end{document}